\documentclass[conferenc]{ieeeconf}
\IEEEoverridecommandlockouts
\usepackage{cite}
\usepackage{amsmath,amssymb,amsfonts}
\usepackage{algorithmic}
\usepackage{lipsum}
\usepackage{multirow}
\usepackage[table]{xcolor} 
\usepackage{graphicx}
\usepackage[caption=false]{subfig}
\usepackage{textcomp}
\usepackage{xcolor}
\usepackage{array}
\usepackage{booktabs}
\usepackage{makecell}
\usepackage{hhline}
\usepackage{soul}
\usepackage{capt-of}
\definecolor{rojo-anaranjado}{HTML}{FF5733}
\definecolor{azul-claro}{HTML}{AED6F1} 
\definecolor{azul-muy-claro}{HTML}{EAF6FF} 
\def\BibTeX{{\rm B\kern-.05em{\sc i\kern-.025em b}\kern-.08em
    T\kern-.1667em\lower.7ex\hbox{E}\kern-.125emX}}
\begin{document}


\title{\LARGE \bf
Behavioural gap assessment of human-vehicle interaction in real and virtual reality-based scenarios in autonomous driving}





\author{S. Martín Serrano$^{*1}$, R. Izquierdo$^{1}$,  I. García Daza$^{1}$, M. A. Sotelo$^1$, D. Fern\'andez-Llorca$^{*1,2}$
\thanks{$^{1}$Computer Engineering Department, Universidad de Alcal\'a, Alcal\'a de Henares, Spain.
}%
        \newline
\thanks{$^{2}$European Commission, Joint Research Centre, Seville, Spain.     
}
\thanks{$^{*}$Corresponding authors: {\tt\small sergio.martin@uah.es,  david.fernandez-llorca@ec.europa.eu}}
}

\maketitle

\begin{abstract}
In the field of autonomous driving research, the use of immersive virtual reality (VR) techniques is widespread to enable a variety of studies under safe and controlled conditions. However, this methodology is only valid and consistent if the conduct of participants in the simulated setting mirrors their actions in an actual environment. 
In this paper, we present a first and innovative approach to evaluating what we term the \emph{behavioural gap}, a concept that captures the disparity in a participant's conduct when engaging in a VR experiment compared to an equivalent real-world situation. To this end, we developed a digital twin of a pre-existed crosswalk and carried out a field experiment (N=18) to investigate pedestrian-autonomous vehicle interaction in both real and simulated driving conditions. In the experiment, the pedestrian attempts to cross the road in the presence of different driving styles and an external Human-Machine Interface (eHMI). 
By combining survey-based and behavioural analysis methodologies, we develop a quantitative approach to empirically assess the \emph{behavioural gap}, as a mechanism to validate data obtained from real subjects interacting in a simulated VR-based environment. 
Results show that participants are more cautious and curious in VR, affecting their speed and decisions, and that VR interfaces significantly influence their actions.


\end{abstract}


\section{Introduction}
As autonomous vehicle (AV) technology advances, the need for rapid prototyping and extensive testing is becoming increasingly important, as real driving tests alone are not sufficient to demonstrate safety \cite{Kalra2016, Llorca2021}. The use of physics-based simulations allows the study of various scenarios and conditions at a fraction of the cost and risk of physical prototype testing, providing valuable insights into the behaviour and performance of AVs in a controlled environment \cite{Schwarz2022}. 

However, one of the main challenges in the development of autonomous driving digital twins is the lack of realism of simulated sensor data and physical models. The so-called \emph{reality gap} can lead to inaccuracies because the virtual world does not adequately generalise all the variations and complexities of the real world \cite{Stocco2022}, \cite{GarciaDaza2023}. Furthermore, despite attempts to generate realistic synthetic behaviours of other road agents (e.g., vehicles, pedestrians, cyclists), simulation lacks empirical knowledge about their behaviour, which negatively affects the gap in behaviour and motion prediction, communication, and human-vehicle interaction \cite{Eady2019}.


Including behaviours and interactions from real agents in simulators is one way to reduce the \emph{reality gap} of autonomous driving digital twins. This can be addressed by using real-time immersive VR \cite{CarlaCHIRA2022, CarlaCHIRA23}. The immersive integration of real subjects into digital twins allows, on the one hand, human-vehicle interaction studies in fully controlled and safe environments. Various HMI modalities can be included to explore extreme scenarios without risk to people and vehicle prototypes. On the other hand, it makes it possible to obtain synthetic sequences from multiple viewpoints (i.e., simulated sensors of AVs) based on the behaviour of real subjects, which can be used to train and test predictive perception models. However, this approach would only be valid if the behaviour of the subjects in the simulated environment is equivalent to their behaviour in a real environment. We refer to this difference in behaviour as the \emph{behavioural gap}, and in order to model it, it is necessary to empirically assess the behaviour of subjects under equivalent real and simulated conditions. 

\begin{figure*}
\centerline{\includegraphics[width=1.95\columnwidth]{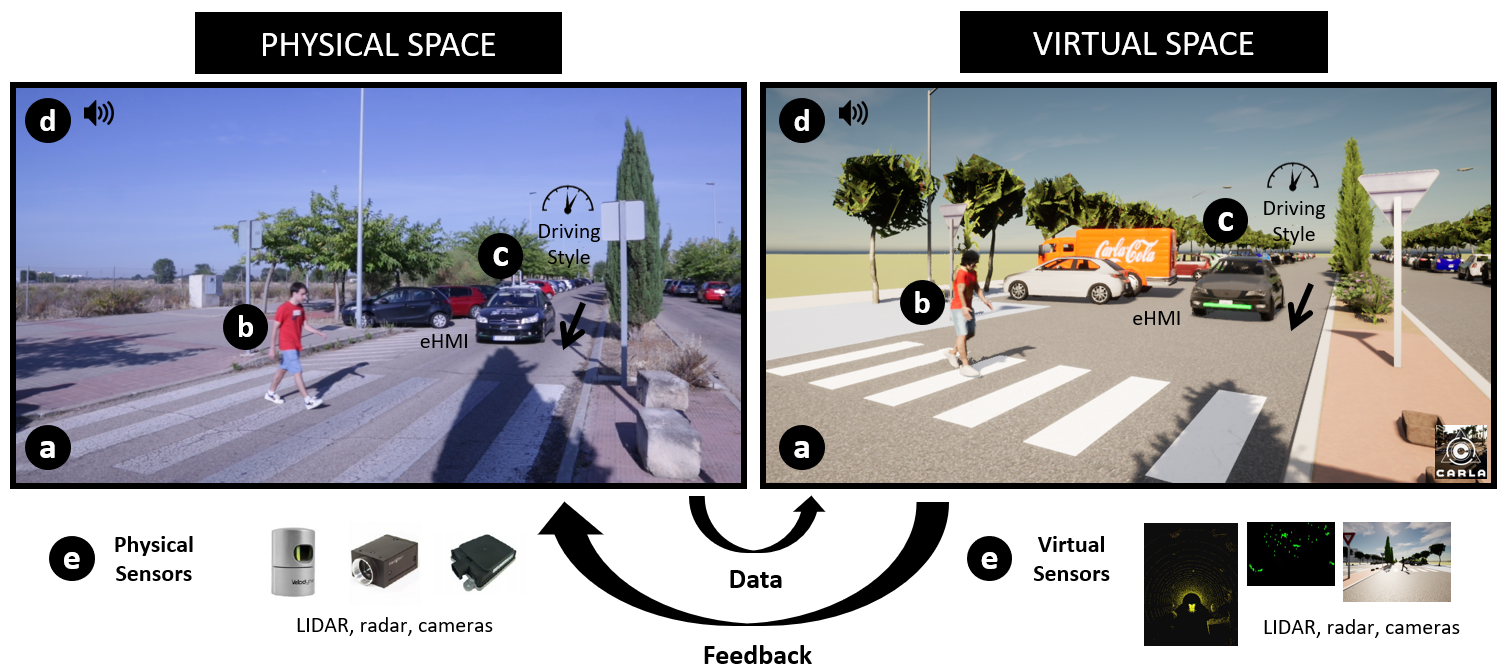}}
\caption{Digital twin for human-vehicle interaction in autonomous driving. (a) 3D crosswalk scenario. (b) Pedestrian attempting to cross. (c) Autonomous vehicle (eHMI, driving style). (d) Ambient sound, lighting and traffic signals. (e) Physical versus virtual sensors.}
\label{fig:schematic1}
\end{figure*}


Meanwhile, the attempt to introduce autonomous driving into daily life makes it crucial to study humans-AVs interactions, as the absence of a driver has an impact on the perception of risk, trust \cite{Li2019} and the level of acceptance by all users \cite{Detjen2021}, including non-driving passengers and external road agents (i.e., pedestrians, cyclists and other drivers) \cite{Llorca2021}. AVs are faced with the need to communicate their intentions using all available resources, which translates into the use of HMIs as a form of explicit communication. Nonetheless, some previous studies suggest the primary basis for crossing decisions taking by pedestrians is the implicit interaction, such us perceived vehicle speeds or safety gap sizes \cite{Clamann2017, Zimmermann2017}. Thus, the first interest of our research is to evaluate together an explicit form of communication (eHMI) and an implicit one, as in this case a different braking manoeuvre of the vehicle. In this paper, we present the results of the first part of our field study on human-AVs interactions, in a real-world crosswalk scenario \cite{Izquierdo2023} and which answers our first research question:

\begin{itemize}
    \item \textbf{RQ1}: \textit{To what extent do the variables "eHMI" and "braking manoeuvre" influence the crossing behaviour of a pedestrian in a real-world crosswalk in terms of (1) vehicle-gazing time, (2) space gap, (3) body-motion, and (4) subjective perception?}
\end{itemize}

On the other hand, we employed a novel framework to insert real agents into the CARLA simulator \cite{CarlaCHIRA2022, CarlaCHIRA23}. Through the CARLA tools and the added motion capture system, we enable an immersive VR interface for a pedestrian and reproduce the same interaction conditions with the vehicle (i.e., eHMI and driving style) \cite{MartinSerrano2023}, allowing us to pose our second research question:

\begin{itemize}
    \item \textbf{RQ2}: \textit{To what extent do the variables "eHMI" and "braking manoeuvre" influence the crossing behaviour of a pedestrian in a virtual crosswalk in terms of (1) vehicle-gazing time, (2) space gap, (3) body-motion, and (4) subjective perception?} 
\end{itemize}

Furthermore, as our interest is focused on providing a pioneering measure of the \textit{behavioural gap} that exists in the activity of a participant depending on whether s/he acts in a physical-real or virtual environment, we developed a digital twin of the exact same crosswalk of the first part of the study, imitating its visibility conditions and road dimensions. The same experiment setup is repeated in a real-world and an identical virtual scenario to answer our last research question:

\begin{itemize}
    \item \textbf{RQ3}: \textit{To what extent does pedestrian crossing behaviour differ between a real and a virtual environment in terms of (1) vehicle-gazing time, (2) space gap, (3) body-motion, and (4) subjective perception?}
\end{itemize}

To our knowledge, this is the first approach that is concerned with evaluating whether human behaviour is realistic within a VR setup for autonomous driving. 


\section{Related work}

\subsection{Understanding Pedestrian-AVs Interaction}

The research of the interactions between pedestrians and AVs is essential to ensure the safety and public acceptance of this emerging technology \cite{Dey2018, Llorca2023}. To date, numerous studies have been conducted to investigate the role of eHMIs and AV driving styles on the pedestrian crossing experience, in controlled real-world environments \cite{DeyMatviien2021, Izquierdo2023} and in VR environments \cite{Nascimento2019, Stadler2019, MartinSerrano2023}. 

Among the eHMI forms commonly explored, we can find several lighting signals designs, textual messages, inclusion of anthropomorphic featuring or trajectory projection on the ground \cite{bazilinskyy2019survey, Furuya2021, mason2022lighting}. For instance, an AV equipped with robotic eyes that look at the pedestrian or head-on helps them make more efficient crossing choices \cite{Chang2022, chang2017eyes}. Various approaches have studied the effect of light-based communication in Wizard-of-Oz designs in which automated driving is simulated that appears to be driverless \cite{Hensch2020-1, Hensch2020-2}. Despite the fact that in many cases visual messages can be displayed on an external surface to indicate the status of the vehicle (e.g., real-time predicted risk levels \cite{Song2023} or directional information \cite{Bazilinskyy2022}), some studies note that their participants prefer direct written instructions to cross the road (i.e., "walk" or "stop") \cite{Ackermann2019, Shuchisnigdha2020-2}. This could be misleading when the traffic situation involves more than one pedestrian \cite{Song2023} so road projection-based eHMIs may be an alternative for scalability to communicate vehicle intentions in shared spaces \cite{Dey2021projections, Mason2022LightingAP, Nguyen2019projections}. Most of the research on eHMI development in virtual reality focuses on visual components, as commercially available hardware and software are at an early stage of development, which poses difficulties in creating multimodal experiences \cite{Le2020}. Auditory, haptic and interactive elements, such as the movement of participants and the virtual representation of their bodies, are mainly used to increase the sense of presence in the virtual environment. However, these elements could also enhance the authenticity of participants' reactions.

In another sense, it has also been shown that pedestrians use implicit communication signals to estimate the behaviour of the vehicle, and apply it to their decisions \cite{Tian2023}. Moreover, leading works suggest that implicit information (i.e., their movement) may be sufficient \cite{Moore2019} or that eHMIs only help convince pedestrians to cross the road when the vehicle speed is ambiguous \cite{DeyMatviien2021}. Deceleration or the distance to the vehicle are more useful in interpreting the intention to yield than the drivers' presence and apparent attentiveness \cite{velasco2021will}. This type of communication has been found to be even more relevant in unmarked locations \cite{Leeknematics,Kalantari2023WhoGF}. 

Although survey-based studies to assess human behaviour in traffic scenes are prevalent \cite{Fridman2017ToWO, Li2018, MERAT2018244}, they fail to collect immediate feedback from experiments \cite{Dijksterhuis2015}. Recording-based studies allow direct measurements and help mitigate potential biases associated with self-reporting \cite{TOM20111794}. Metrics extracted from objective data can be treated as dependent variables and analysed using a linear mixed models, including road-crossing decision times, gaze-based times, crossing speed or distances to the vehicle \cite{FengYan2023, Guo2022}.

\subsection{Bridging the Simulation-to-Reality Gap}

Testing in simulated environments offers some advantages over real-world testing, such as more safety for participants in the experiments and the facility of constructing scenarios \cite{mti7020016}. This saves a lot of costs in terms of time and effort. However, differences in lighting, textures, vehicle dynamics and agents behaviour between simulated and real environments raise doubts about the validity of the results in this new context \cite{Hu2024}.

The first approach to assessing whether simulation-based testing can be a reliable substitute for real-world testing is to validate the virtual models of the sensors by determining whether their discrepancy with reality is sufficiently low. We found works that do this in the case of radar \cite{Anthony2021} and a camera-based object detection algorithm \cite{Reway2020}. Typically, the gap between synthetic and real-world datasets is well-known \cite{Gadipudi2022}, and there are already proposals to alleviate it as methods that obtain realistic images from those recorded in simulation or that bridge the differences in system dynamics \cite{Cruz2020,icinco22}. We emphasise that the gap worsens in multi-agent systems due to the complexity of transferring agent interactions and the synchronisation of the environment \cite{Candela2022}.

One of the strategies researchers employ to bridge the gap between simulation and reality in autonomous driving are the digital twins (DTs) \cite{Hu2024, Yu2022, Almeaibed2021, Ge2019, Yun2021}. Some study utilises a real small-scale physical vehicle and its digital twin to investigate the transferability of behaviour and failure exposure between virtual and real-world environments \cite{Stocco2022}. There have been no previous approaches to assess the gap in the behaviour of real agents (e.g., pedestrians) within a simulation, as we do in this work with a full-scale digital twin of a scenario and immersive VR for real-time interaction with an AV.



\section{Method}
\subsection{Experiment Design}

The currently study presents improvements over previous immersive VR experiments with pedestrians, since \emph{(i)} it is conducted in the CARLA simulator \cite{carla2017} and not in Unity, which allows the use of highly specialised functions for autonomous driving, and \emph{(ii)} a motion capture system is added to accurately collect the participants motion data. On one hand, we can assess interactions by the usual methods, such as eye contact with the vehicle or questionnaires \cite{Rasouli2020}. Furthermore, we combine explicit and implicit communication under safe conditions, and capture the behaviour of the participants by video and inertial sensors.

\subsubsection{Experiment Scenario Design}

An existing crosswalk on the area of the University of Alcalá (Spain), was chosen to perform the real driving tests and also as the baseline to construct the VR environment (see Fig. \ref{fig:schematic1}). In this scenario, an AV drives on a day with plenty of sunlight along a street in a straight line until it reaches a crosswalk. The pedestrian, who wishes to cross the road perpendicularly, needs to take 2-3 steps to have visibility to their left side (due to other parked vehicles and vegetation).

The map model is downloaded from OpenStreetMap \cite{Open2024} and converted to a Unreal Engine project where the elements are detailed. From the vehicle blueprints offered by CARLA we choose the model and colour of the physical vehicle and attach the sensors to perceive its surroundings (i.e., LiDAR, radar and cameras).

\begin{figure}
\centerline{\includegraphics[width=\columnwidth]{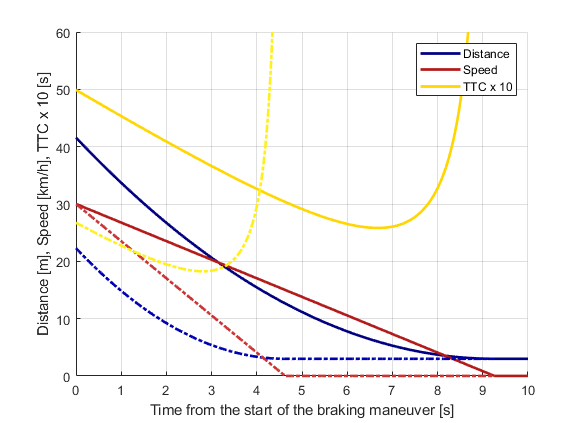}}
\caption{Smooth (continuous line) and aggressive (dashed line) deceleration.}
\label{fig:varibles_at_crossing_gentle}
\end{figure}

In order to facilitate interaction, the pedestrian waits with their back to the road and is instructed to turn around when the vehicle is at a distance of about 40 meters. As can be seen in Fig. \ref{fig:varibles_at_crossing_gentle}, two braking manoeuvres were designed. In both cases, the vehicle travels at a speed of 30 km/h and applies a constant deceleration of -0.9 m/s\textsuperscript{2} (smooth) or -1.8 m/s\textsuperscript{2} (aggressive) until it comes to a complete stop in front of the crosswalk and yields the right-of-way. This is done to study whether the pedestrian perceives the situation as more risky when the vehicle brakes with less anticipation and the time-to-collision (TTC) is smaller. 

\begin{figure}
\centering
\subfloat[]{\includegraphics[width=0.48\columnwidth]{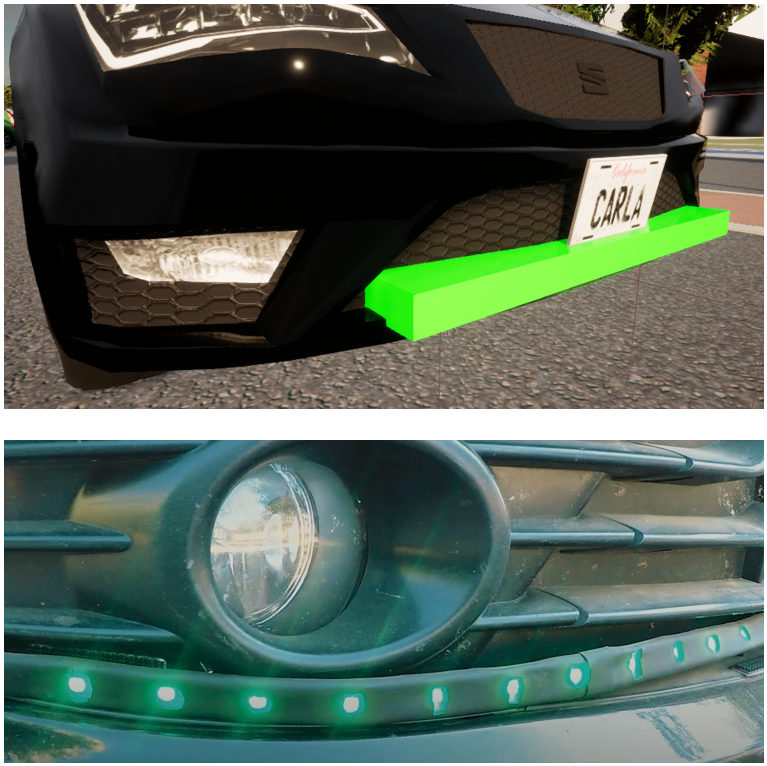}%
\label{fig:grail_green}}
\hfil
\subfloat[]{\includegraphics[width=0.48\columnwidth]{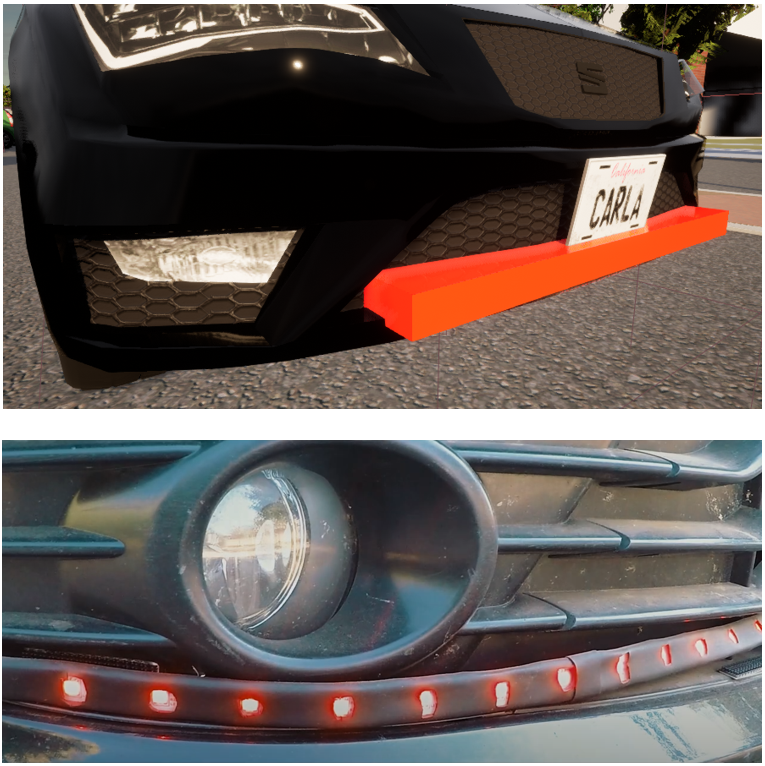}%
\label{fig:grail_red}}
\caption{AV with eHMI activated communicating green (a) and red (b) status. Virtual (above) and physical (bottom) design. }
\label{fig:external_HMI}
\end{figure}

To alert the pedestrian of its intention to yield, the vehicle was equipped with the GRAIL (Green Assistant Interfacing Light)  system \cite{GRAIL}. As shown in Figs. \ref{fig:grail_green} and \ref{fig:grail_red}, the AV uses green to indicate awareness of the pedestrian (which implies that it will stop if necessary), and red to warn that nothing prevents it from continuing on its way. It is also possible that the interface is deactivated so the pedestrian does not have any explicit information about the vehicle intention. This front-end design is sufficient for the specific scenario of this work. However, more complex scenarios with poorer visibility conditions might require enhancements, such as extending the LED light-band to the sides of the vehicle, or even incorporating a 360-degree eHMI approach~\cite{Hub2023}. 

\subsubsection{Experiment Task Design}

The combination of eHMI on or off, and the different strategies of deceleration result in the road-crossing tasks listed in Table \ref{tab:testsetup}. When activated, the eHMI starts emitting the red light and changes to green when the vehicle has covered a 30 \% of the braking distance (12 or 6 meters, depending on the type of manoeuvre). All tasks were performed in a random order specific to each participant, except for task 0 (warm-up task), which always started the experiment and in which the vehicle did not stop and the participant only had to turn towards the road and watch the vehicle without initiating the crossing action.

\begin{table}[t]
\renewcommand{\arraystretch}{1.1}
\caption{Experimentation Tasks settings}
\begin{center}
\begin{tabular}{c|c|c|c|c}
\textbf{Task}& \textbf{AV Strategy} & \textbf{External} & \textbf{Stop}\\
\textbf{Number}& \textbf{Deceleration} & \textbf{HMI} & \\
\hline
0   & -             & -    & No \\
1   & Smooth        & -     & Yes\\
2   & Aggressive    & -     & Yes\\
3   & Smooth        & Activated & Yes\\
4   & Aggressive    & Activated & Yes\\
\end{tabular}
\label{tab:testsetup}
\end{center}
\end{table}

\subsection{Virtual Reality Apparatus}

Tests under simulated driving conditions were conducted in a VR space of 8 meters long x 3 meters wide. The virtual environment was constructed under a 1:1 scheme mapped to the real-life environment, so participants adopted the real-walking locomotion style, leading to a more realistic movement and a greater sense of presence. 

We use a specific framework for the insertion of real agents in CARLA \cite{CarlaCHIRA2022, CarlaCHIRA23}. An immersive interface is enabled in VR for the incorporation of a pedestrian into the traffic scene. Some of the features added to the simulator were real-time avatar control, positional sound or body tracking. The Meta Quest 2 headset was connected via WiFi to a Windows 10 desktop and an NVIDIA GeForce RTX 3060 graphics card. We chose Perception Neuron Studio \cite{PNS2022} as the motion capture system to record the user's pose and integrate it into the scenario.

\subsection{Experiment Procedure}

The experimental procedure differed between the real and virtual contexts, yet it could be distinctly delineated into four phases:


\begin{enumerate}
    \item \textit{Introduction}: At the beginning, participants were provided with written information about the experiment, such as its purpose, the explanation of the AV and the functionality of the eHMI. They were also assigned a unique anonymous identifier and were assured of their ability to discontinue the experiment at any time if they so desired. Lastly, they were asked to sign the consent to participate as subjects in the study. 
    \item \textit{Familiarisation (warm-up)}: Participants were aided in donning the inertial sensors and VR headset, following which they were invited to explore the virtual environment void of any vehicular traffic. Subsequently, the Perception Neuron system underwent calibration, and the initial task of the experiment (task 0) was conducted as an illustrative example.
    \item \textit{Experimentation}: Throughout this phase, participants completed the four tasks of the experiment (see Table \ref{tab:testsetup}) while answering questions posed by an accompanying researcher about their subjective perception.
    \item \textit{Filling in the post-questionnaire}: After concluding the experiment, participants removed the VR headset and inertial sensors and, in both the real and virtual context, were asked to fill out a post-questionnaire.
\end{enumerate}

\subsection{Data Collection}

During the experiment various types of data were collected to analyse the resulting pedestrian-AV interactions, including objective measurements (i.e., movement path, gaze time) as well as responses to questionnaires.

In the first instance, the AV in real configuration was fitted with a RTK-GPS system that provided its precise position with respect to the crosswalk and served as a reference for applying the braking manoeuvre, while an external camera mounted on the top of the vehicle recorded the environment at 10 Hz. Within Unreal Engine 4 and Axis Studio \cite{CarlaCHIRA2022, CarlaCHIRA23}, all data from the VR experiment were recorded as the (1) timestamp, (2) vehicle's position and parameters (i.e., coordinate x, y, z, rotation, brake, steer, throttle, gear), (3) participant's position and animation (i.e., coordinate x, y, z, rotation, .fbx), and (4) playbacks of the Quest 2 view, the VR setup, and from within the simulator, synchronised at 18.8 Hz.

The questionnaire collected participant's information (e.g., age, gender, familiarity with AVs and with VR) and subjective feedback on the influence of the different types of communication in each interaction through the following questions: 

\begin{enumerate}
\item[Q1:] How safe did you feel at the scene? 
\item[Q2:] How aggressive did you perceive the braking manoeuvre of the vehicle?
\item[Q3:] Did the visual communication interface improve your confidence to cross? 
\end{enumerate}

Answers were tabulated on a 7-step Likert scale \cite{joshi2015likert}. In addition, the participants completed a 15-item presence scale (depicted in Appendix A) to evaluate the quality of pedestrian immersion in the scene.

\subsection{Participant’s Characteristics}

A total of 18 participants, aged between 24 and 62 years (M = 40.11, SD = 11.62), with a gender distribution of 33$\%$ women and 67$\%$ men, were recruited from both inside and outside the university area and engaged in the experiment. 

In regard to familiarity with AVs, 38.9$\%$ had extensive knowledge of the subject, another 38.9$\%$ considered that they had an average knowledge of Advanced Driver Assistance Systems (ADAS), 44.4$\%$ had previously interacted with an AV (either as a user or pedestrian) compared to 55.6$\%$ who had not, and 22.2$\%$ had no prior exposure or understanding of AVs. For VR experience, the majority of participants had either never used VR goggles (50$\%$) or had only tried them once before (33.3$\%$). All participants had normal vision or wore corrective glasses (22.2$\%$) that they kept when fitting the VR headset, and had normal mobility, so they were able to complete the experiment successfully.

\subsection{Data Analysis}

Different metrics can be acquired from the objective data (i.e., movement trajectory, gaze point) gathered during the experiments. The metrics chosen for analysis in this research are defined as follows:

\begin{itemize}
    \item[\textbullet] \emph{Vehicle-gazing time while crossing ($T_{c}$):} it represents the cumulative duration of gazing at the AV while crossing, as inferred from the collected eye gazing data.
    \item[\textbullet] \emph{Crossing initiation time (CIT):} 
    computed as the interval from when the pedestrian visually identifies the AV until s/he decides to cross. If the pedestrian crosses before noticing the AV, then CIT is zero.
    \item[\textbullet] \emph{Vehicle-gazing time ($T_{av}$):} it represents the cumulative duration of gazing at the AV throughout the entire crossing process, as inferred from the collected eye gazing data. That is, $T_{av} = CIT + T_c$.     
    \item[\textbullet] \emph{Space gap (L):} the distance between the AV and the pedestrian, measured from the AV to the centre of the crosswalk when the pedestrian decides to cross.
    \item[\textbullet] \emph{Pace cadence ($F_{p}$):} defined as the dominant step frequency at which the pedestrian crosses the road.
    \item[\textbullet] \emph{Gait cycles ($G$):} referring to the number of gait cycles when the pedestrian makes the decision to cross, along with the stabilisation times of the two ankles.
\end{itemize}

\begin{figure}
\centering
\renewcommand\thesubfigure{\roman{subfigure}}
\subfloat[]{\includegraphics[width=0.15\textwidth]{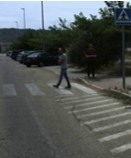}%
\label{fig_ped_1}}
\hfil
\subfloat[]{\includegraphics[width=0.15\textwidth]{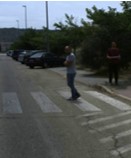}%
\label{fig_ped_2}}
\hfil
\subfloat[]{\includegraphics[width=0.15\textwidth]{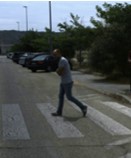}%
\label{fig_ped_3}}
\caption{Crossing decision event. (i) The pedestrian takes two steps forward to gain visibility. (ii) The vehicle is approaching and the pedestrian slows down without stopping. (iii) The pedestrian makes the decision to cross. }
\label{fig:decision_crossing_event}
\end{figure}

To obtain the above indicators, the crossing intention event is defined as the moment the pedestrian decides to cross the crosswalk and is extracted from the video recordings and the reconstructed trajectory in the virtual environment. The rules for identifying the event state are the following:

\begin{enumerate}
    \item In case the pedestrian is stopped and starts to move into the crosswalk, the decision is made at the first frame in which the movement is discernible.
    \item If there is not a stop and the pace is slowed, the decision occurs at the frame the pedestrian starts accelerating.
    \item If there is no alteration in the pedestrian's speed, the decision is made upon sighting the vehicle.
    \item If the pedestrian does not look at the vehicle, we take the first frame when the pedestrian appears on the vehicle's front camera.
\end{enumerate} 

An example of the crossing decision in the real environment can be seen in Fig. \ref{fig:decision_crossing_event}.

Ultimately, to conclusively state that there are differences in crossing decision making in each task of the experiment, we employed the Student's t-test \cite{student}. For the analysis of the questionnaire, we used the Wilcoxon signed-rank test \cite{woolson2007wilcoxon}.

\section{Results}

\begin{figure*}[t]
\centering
\includegraphics[width=1.0\textwidth]{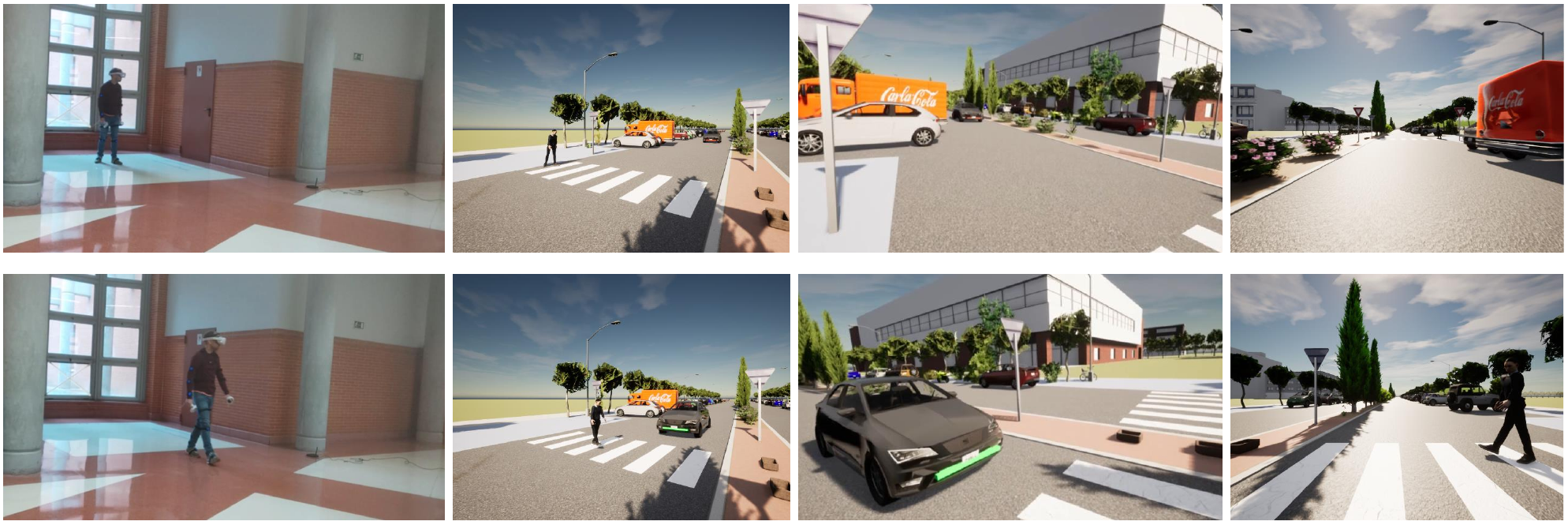}
\caption{Pedestrian-AV interaction in VR setup. (Upper row) The pedestrian waits while eHMI displays a red status. (Lower row) The eHMI switches to green status and the pedestrian decides to cross. From left to right: VR experimentation environment; overview of the simulated virtual scenario; pedestrian perspective; AV perspective (simulated camera).}
\label{fig:example_interacion_exterior}
\end{figure*}

This section presents the results obtained in the experiment with the real and virtual setup. As can be seen in Fig. \ref{fig:example_interacion_exterior}, the VR headset projects the crosswalk onto its lenses and allows mobility around the scene. We aim to examine the significant effects of implicit and explicit vehicle communication on pedestrian crossing behaviour.


\subsection{Vehicle Gazing ($T_{av}$) and Crossing Initiation Times (CIT)}

\begin{figure*}[t]
\centering
\includegraphics[width=1.0\textwidth]{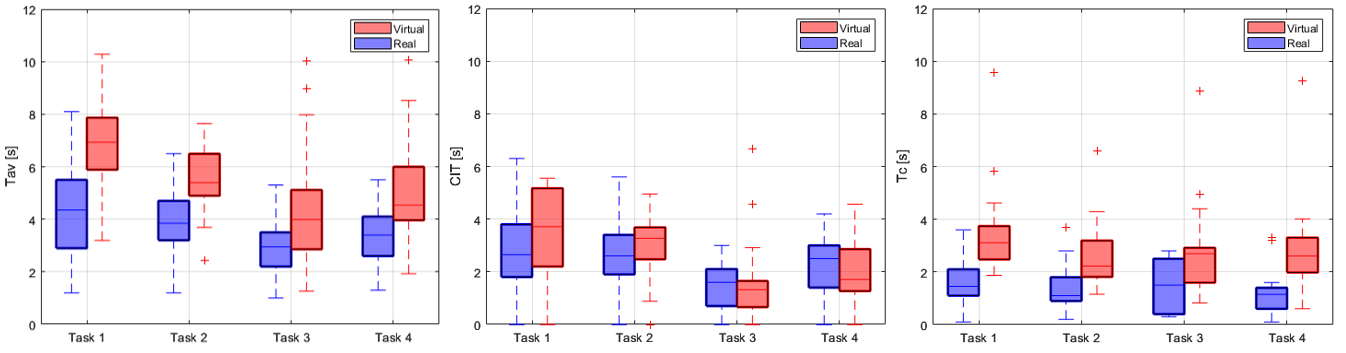}
\caption{Gazing times in the Virtual and Real environment.}
\label{fig:cajas_tiempos}
\end{figure*}

To establish categorical statements about the impact of the braking manoeuvre or eHMI on the crossing decision, we utilise the Student's t-test \cite{student}. The procedure for this test involves calculating the difference between the means of two groups of samples and adjusting this difference for within-group variability and sample size. This adjusted difference is compared to a probability t-distribution to determine if it is large enough to be considered significant. If this happens with the means of the data extracted from the experimental tasks, the null hypothesis ($H_0:\mu_i\leq\mu_j$) is rejected in favour of the alternative hypothesis ($H_1:\mu_i>\mu_j$). Fig. \ref{fig:cajas_tiempos} shows the box-plots of the gaze duration to the vehicle in the tests, considering combinations of two factors: deceleration type and activation of the eHMI (see Table \ref{tab:testsetup} for details).

The Table \ref{tab:student-tiempos} expresses categorical statements, i.e., a 1 in a cell means rejection of the null hypothesis and acceptance of the alternative hypothesis with a confidence level of 95\%, meaning the gaze times in task $i$ (in the row) are significantly larger than those in task $j$ (in the column).

A first aspect to highlight is that the active eHMI decreases the observation times in the two experimental setups ($T_{av}$: t1 $>$ t3 and t2 $>$ t4). This effect cannot be appreciated as directly comparing the two types of deceleration, since the vehicle approaches at different speeds and does not reach the crosswalk at the same time. To analyse the time pedestrians spend observing the vehicle before crossing, we must focus on the CIT, which eHMI shortens by maintaining a smooth deceleration (CIT: t1 $>$ t3). The same impact of eHMI during aggressive deceleration is only seen in the virtual setup (CIT: t2 vs t4).

\begin{table}[htbp]
\renewcommand{\arraystretch}{1.2}
\caption{Gazing times, Student t-test, $\alpha$=0.05}
\begin{center}
\begin{tabular}{cccc|p{0.8cm}p{0.8cm}p{0.8cm}p{0.8cm}}
\multicolumn{4}{c|}{\textbf{$H_1:\mu_i>\mu_j$}}& \multicolumn{4}{c}{\textbf{Task number $j$}} \\
 \multicolumn{4}{c|}{} & \colorbox{gray!0}{1} & \colorbox{gray!0}{2} & \colorbox{gray!0}{3} & \colorbox{gray!0}{4}\\
\hline
\multirow{28.5}{*}{\rotatebox[origin=c]{90}{\textbf{Task number $i$}}} 

& \multirow{9}{*}{\rotatebox[origin=c]{90}{\textbf{Tav}}} & \multirow{4.5}{*}{\rotatebox[origin=c]{90}{\textbf{Real setup}} \newline \rotatebox[origin=c]{90}{\textbf{testing}}}                                                    
\rule{0pt}{9pt}& 1   &\colorbox{gray!0}{--}  &\colorbox{gray!20}{0}  &\colorbox{gray!0}{1}  &\colorbox{gray!0}{1} \\
\rule{0pt}{10pt}&&   &2   &\colorbox{gray!0}{0}  &\colorbox{gray!0}{--}  &\colorbox{gray!0}{1}  &\colorbox{gray!0}{1} \\
\rule{0pt}{10pt}&&  &3   &\colorbox{gray!0}{0}    &\colorbox{gray!0}{0}     &\colorbox{gray!0}{--}     & \colorbox{gray!0}{0}\\
\rule{0pt}{10pt}&&  &4   &\colorbox{gray!0}{0}  &\colorbox{gray!0}{0}  &\colorbox{gray!20}{1}  &\colorbox{gray!0}{--} \\

\cline{4-8}
&  &\multirow{4.5}{*}{\rotatebox[origin=c]{90}{\textbf{Virtual setup}} \newline \rotatebox[origin=c]{90}{\textbf{testing}}}                        

\rule{0pt}{9.5pt}&1   &\colorbox{gray!0}{--}   &\colorbox{gray!20}{1}  &\colorbox{gray!0}{1}  &\colorbox{gray!0}{1}  \\
\rule{0pt}{10pt}&& &2   &\colorbox{gray!0}{0}  &\colorbox{gray!0}{--}   &\colorbox{gray!0}{1}  &\colorbox{gray!0}{1}\\
\rule{0pt}{10pt}&&  &3   &\colorbox{gray!0}{0}     &\colorbox{gray!0}{0}     &\colorbox{gray!0}{--}     &\colorbox{gray!0}{0}      \\
\rule{0pt}{10pt}&&  &4   &\colorbox{gray!0}{0}  &\colorbox{gray!0}{0}    &\colorbox{gray!20}{0}     &\colorbox{gray!0}{--}     \\

\cline{3-8}
& \multirow{9}{*}{\rotatebox[origin=c]{90}{\textbf{CIT}}} & \multirow{4.5}{*}{\rotatebox[origin=c]{90}{\textbf{Real setup}} \newline \rotatebox[origin=c]{90}{\textbf{testing}}}                                                    
\rule{0pt}{10pt}& 1   &\colorbox{gray!0}{--}  &\colorbox{gray!0}{0}  &\colorbox{gray!0}{1}  &\colorbox{gray!0}{1} \\
\rule{0pt}{10pt}&&   &2   &\colorbox{gray!0}{0}  &\colorbox{gray!0}{--}  &\colorbox{gray!0}{1}  &\colorbox{gray!20}{0} \\
\rule{0pt}{10pt}&&  &3   &\colorbox{gray!0}{0}    &\colorbox{gray!0}{0}     &\colorbox{gray!0}{--}     & \colorbox{gray!0}{0}\\
\rule{0pt}{10pt}&&  &4   &\colorbox{gray!0}{0}  &\colorbox{gray!0}{0}  &\colorbox{gray!20}{1}  &\colorbox{gray!0}{--} \\

\cline{4-8}
&  &\multirow{4.5}{*}{\rotatebox[origin=c]{90}{\textbf{Virtual setup}} \newline \rotatebox[origin=c]{90}{\textbf{testing}}}                        

\rule{0pt}{10pt}&1   &\colorbox{gray!0}{--}   &\colorbox{gray!0}{0}  &\colorbox{gray!0}{1}  &\colorbox{gray!0}{1}  \\
\rule{0pt}{10pt}&& &2   &\colorbox{gray!0}{0}  &\colorbox{gray!0}{--}   &\colorbox{gray!0}{1}  &\colorbox{gray!20}{1}\\
\rule{0pt}{10pt}&&  &3   &\colorbox{gray!0}{0}     &\colorbox{gray!0}{0}     &\colorbox{gray!0}{--}     &\colorbox{gray!0}{0}      \\
\rule{0pt}{10pt}&&  &4   &\colorbox{gray!0}{0}  &\colorbox{gray!0}{0}    &\colorbox{gray!20}{0}     &\colorbox{gray!0}{--}     \\

\cline{3-8}
& \multirow{9}{*}{\rotatebox[origin=c]{90}{\textbf{Tc}}} & \multirow{4.5}{*}{\rotatebox[origin=c]{90}{\textbf{Real setup}} \newline \rotatebox[origin=c]{90}{\textbf{testing}}}                                                    
\rule{0pt}{10pt}& 1   &\colorbox{gray!0}{--}  &\colorbox{gray!20}{0}  &\colorbox{gray!20}{0}  &\colorbox{gray!0}{1} \\
\rule{0pt}{10pt}&&   &2   &\colorbox{gray!0}{0}  &\colorbox{gray!0}{--}  &\colorbox{gray!0}{0}  &\colorbox{gray!0}{0} \\
\rule{0pt}{10pt}&&  &3   &\colorbox{gray!0}{0}    &\colorbox{gray!0}{0}     &\colorbox{gray!0}{--}     & \colorbox{gray!0}{0}\\
\rule{0pt}{10pt}&&  &4   &\colorbox{gray!0}{0}  &\colorbox{gray!0}{0}  &\colorbox{gray!0}{0}  &\colorbox{gray!0}{--} \\

\cline{4-8}
&  &\multirow{4.5}{*}{\rotatebox[origin=c]{90}{\textbf{Virtual setup}} \newline \rotatebox[origin=c]{90}{\textbf{testing}}}                        

\rule{0pt}{10pt}&1   &\colorbox{gray!0}{--}   &\colorbox{gray!20}{1}  &\colorbox{gray!20}{1}  &\colorbox{gray!0}{1}  \\
\rule{0pt}{10pt}&& &2   &\colorbox{gray!0}{0}  &\colorbox{gray!0}{--}   &\colorbox{gray!0}{0}  &\colorbox{gray!0}{0}\\
\rule{0pt}{10pt}&&  &3   &\colorbox{gray!0}{0}     &\colorbox{gray!0}{0}     &\colorbox{gray!0}{--}     &\colorbox{gray!0}{0}      \\
\rule{0pt}{10pt}&&  &4   &\colorbox{gray!0}{0}  &\colorbox{gray!0}{0}    &\colorbox{gray!0}{0}     &\colorbox{gray!0}{--}     \\
                
\end{tabular}
\label{tab:student-tiempos}
\end{center}
\end{table}

\begin{table}[htbp]
\renewcommand{\arraystretch}{1.1}
\captionsetup{skip=-20pt}
\caption{Certainty of the discrepancy, Student t-test}
\begin{center}
\begin{tabular}{cc|p{0.8cm}p{0.8cm}p{0.8cm}p{0.8cm}}
 \multicolumn{2}{c|}{Alternative hypothesis:} & \textbf{Task 1} & \textbf{Task 2} & \textbf{Task 3} & \textbf{Task 4}\\
\cline{1-6}
\multirow{2.25}{*}{\rotatebox[origin=c]{90}{\textbf{Tav}} }

    \rule{-10pt}{10pt}&$\mu_{virtual}>\mu_{real}$ & \colorbox{gray!20}{99.9 \%}& \colorbox{gray!20}{99.9 \%}& \colorbox{gray!20}{99.4  \%}& \colorbox{gray!20}{99.7 \%} \\
    \rule{-10pt}{0pt}&$\mu_{real}>\mu_{virtual}$ & \colorbox{gray!0}{0.0 \%}& \colorbox{gray!0}{0.0 \%}& \colorbox{gray!0}{0.6 \%}& \colorbox{gray!0}{0.3 \%} \\ 
    
\cline{2-6}
\multirow{2.25}{*}{\rotatebox[origin=c]{90}{\textbf{CIT}} }

    \rule{-10pt}{10pt}&$\mu_{virtual}>\mu_{real}$ & \colorbox{gray!20}{95.9 \%}& \colorbox{gray!0}{88.4 \%}& \colorbox{gray!0}{67.5 \%}& \colorbox{gray!0}{33.9 \%} \\
    \rule{-10pt}{0pt}&$\mu_{real}>\mu_{virtual}$ & \colorbox{gray!0}{4.1 \%}& \colorbox{gray!0}{11.6 \%}& \colorbox{gray!0}{32.5 \%}& \colorbox{gray!0}{66.1 \%} \\   
    
\cline{2-6}
\multirow{2.25}{*}{\rotatebox[origin=c]{90}{\textbf{Tc}} }

    \rule{-10pt}{10pt}&$\mu_{virtual}>\mu_{real}$ & \colorbox{gray!20}{99.9 \%}& \colorbox{gray!20}{99.9 \%}& \colorbox{gray!20}{99.7  \%}& \colorbox{gray!20}{99.9 \%} \\
    \rule{-10pt}{0pt}&$\mu_{real}>\mu_{virtual}$ & \colorbox{gray!0}{0.0 \%}& \colorbox{gray!0}{0.0 \%}& \colorbox{gray!0}{0.3 \%}& \colorbox{gray!0}{0.0 \%} \\

\end{tabular}
\label{tab:tiempos_gap}
\end{center}
\end{table}

When comparing the two scenarios, rather than making categorical statements, we show the probability of significant discrepancy between the tasks that will be used at the end of the research to quantify the \emph{behavioural gap}. It is evident from the data provided in Table \ref{tab:tiempos_gap} that the gaze duration is greater in the virtual environment. Upon separately examining the time preceding and following the decision to cross, we see that the disparity is less pronounced within the CIT. In the absence of eHMI, pedestrians observe more of the vehicle before crossing in the virtual setup (CIT: t1$_{virtual}$ $>$ t1$_{real}$ and t2$_{virtual}$ vs t2$_{real}$) while, if eHMI is activated, the CIT resembles more closely. The notable differences in vehicle gazing times between both setups and across all experiment variations occur while walking on the road (T$_{c}$: t$_{i,virtual}$ $>$ t$_{i,real}$). This suggests that pedestrians pay significantly more attention to the vehicle after making the decision to cross when they are interacting in the virtual environment.

\subsection{Space Gap (L)}

\begin{figure}[t]
\centerline{\includegraphics[width=\columnwidth]{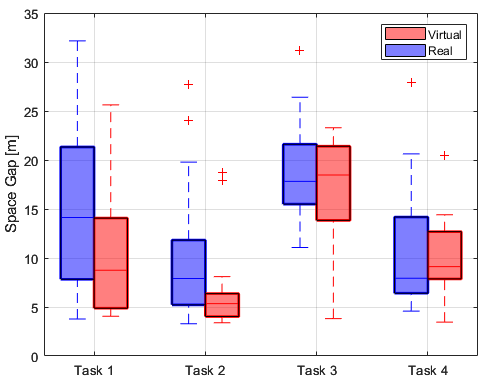}}
\caption{Box-plots of the pedestrian-AV distances at the crossing decision. Virtual and Real testing.}
\label{fig:boxplot_distance}
\end{figure}

Box-plots of the space gap in each task (i.e., the distance between the pedestrian and the AV at the crossing decision) are depicted in Fig. \ref{fig:boxplot_distance} for both real and virtual environments. In addition, Table \ref{tab:student} shows the results of the Student's t-test to evaluate the impact of eHMI and the type of manoeuvre.

\begin{table}[htbp]
\renewcommand{\arraystretch}{1.2}
\caption{Space Gap, Student t-test, $\alpha$=0.05}
\begin{center}
\begin{tabular}{ccc|p{0.9cm}p{0.9cm}p{0.9cm}p{0.9cm}}
\multicolumn{3}{c|}{\textbf{$H_1:\mu_i>\mu_j$}}& \multicolumn{4}{c}{\textbf{Task number $j$}} \\
 \multicolumn{3}{c|}{} & \colorbox{gray!0}{1} & \colorbox{gray!0}{2} & \colorbox{gray!0}{3} & \colorbox{gray!0}{4}\\
\hline
\multirow{9.5}{*}{\rotatebox[origin=c]{90}{\textbf{Task number $i$}}} & \multirow{4.5}{*}{\rotatebox[origin=c]{90}{\textbf{Real setup}} \newline \rotatebox[origin=c]{90}{\textbf{testing}}}                                                   
\rule{0pt}{9pt}& 1   &\colorbox{gray!0}{--}  &\colorbox{gray!0}{1}  &\colorbox{gray!0}{0}  &\colorbox{gray!20}{1} \\
\rule{0pt}{10pt}&&2   &\colorbox{gray!0}{0}  &\colorbox{gray!0}{--}  &\colorbox{gray!0}{0}  &\colorbox{gray!0}{0} \\
\rule{0pt}{10pt}&&3   &\colorbox{gray!0}{1}    &\colorbox{gray!0}{1}     &\colorbox{gray!0}{--}     & \colorbox{gray!0}{1}\\
\rule{0pt}{10pt}&&4   &\colorbox{gray!0}{0}  &\colorbox{gray!20}{0}  &\colorbox{gray!0}{0}  &\colorbox{gray!0}{--} \\

\cline{3-7}
&\multirow{5.5}{*}{\rotatebox[origin=c]{90}{\textbf{Virtual setup}} \newline \rotatebox[origin=c]{90}{\textbf{testing}}}                        

\rule{0pt}{9.5pt}& 1   &\colorbox{gray!0}{--}   &\colorbox{gray!0}{1}  &\colorbox{gray!0}{0}  &\colorbox{gray!20}{0}  \\
\rule{0pt}{10pt}&&2   &\colorbox{gray!0}{0}  &\colorbox{gray!0}{--}   &\colorbox{gray!0}{0}  &\colorbox{gray!0}{0}\\
\rule{0pt}{10pt}&&3   &\colorbox{gray!0}{1}     &\colorbox{gray!0}{1}     &\colorbox{gray!0}{--}     &\colorbox{gray!0}{1}      \\
\rule{0pt}{10pt}&&4   &\colorbox{gray!0}{0}  &\colorbox{gray!20}{1}    &\colorbox{gray!0}{0}     &\colorbox{gray!0}{--}     \\
                
\end{tabular}
\label{tab:student}
\end{center}
\end{table}

With a confidence level of 95\%, we assert that the smooth braking manoeuvre increases the distance to the vehicle when pedestrians decide to cross (Space gap: t1 $>$ t2 and t3 $>$ t4). The same applies to eHMI activation while maintaining the smooth braking manoeuvre (Space gap: t3 $>$ t1). However, although the impact of the eHMI persists in the virtual setup by maintaining aggressive braking, this is not the case in the real setup (Space gap: t4 vs t2).

\begin{table}[tbp]
\renewcommand{\arraystretch}{1.1}
\captionsetup{skip=-20pt}
\caption{Certainty of the discrepancy in Space Gap, Student t-test}
\begin{center}
\begin{tabular}{cc|p{0.8cm}p{0.8cm}p{0.8cm}p{0.8cm}}
 \multicolumn{2}{c|}{Alternative hypothesis:} & \textbf{Task 1} & \textbf{Task 2} & \textbf{Task 3} & \textbf{Task 4}\\
\cline{2-6}
\multirow{6.5}{*}{}

    \rule{-15pt}{10pt}&$\mu_{virtual}>\mu_{real}$ & \colorbox{gray!0}{0.1 \%}& \colorbox{gray!0}{0.6 \%}& \colorbox{gray!0}{12.0 \%}& \colorbox{gray!0}{22.4 \%} \\
    \rule{-15pt}{0pt}&$\mu_{real}>\mu_{virtual}$ & \colorbox{gray!20}{99.8 \%}& \colorbox{gray!20}{99.3 \%}& \colorbox{gray!0}{88.0 \%}& \colorbox{gray!0}{77.6 \%} \\

\end{tabular}
\label{tab:distances_gap}
\end{center}
\end{table}

The Table \ref{tab:distances_gap} provides a direct comparison of space gaps across both setups. The findings indicate that the participants cross significantly earlier in the real setup (i.e., with a larger space gap) when the eHMI is deactivated (Space Gap: t1$_{real}$ $>$ t1$_{virtual}$ and t2$_{real}$ $>$ t2$_{virtual}$). Concerning experimental tasks which employ explicit communication (t3 and t4), the values of space gap exhibit greater similarity, leading to the non-rejection of the null hypothesis and, thus, precluding any definitive statement.

\subsection{Body Motion}

Among the advantages of inserting real agents into a simulation environment \cite{CarlaCHIRA2022, CarlaCHIRA23} is the possibility of generating synthetic sequences from various perspectives and configurations. To accomplish this, it is necessary to reconstruct the trajectory and 3D pose of the participant within the scenario, for which Perception Neuron's sensors and software provide an .fbx file over time \cite{PNS2022}. This approach allows an accurate analysis of the participant's motion style throughout the experiments.

\begin{figure} 
\begin{minipage}{5.5cm} 
\begin{center}
\scriptsize
\begin{tabular}{p{1.6cm}p{1.2cm}p{1.2cm}} 
\rule{0pt}{10pt}Right - Left & \rule{-9.5pt}{10pt}DWPose & \rule{-7.5pt}{10pt}PNs \\ \hline
Trunk & \rule{-12pt}{10pt}100.0 \% & \rule{-10pt}{10pt}100.0 \% \\
Neck & \rule{-12pt}{10pt}33.9 \% & \rule{-10pt}{10pt}24.9 \% \\
Shoulder & \rule{-12pt}{10pt}31.1 - 31.1 \% & \rule{-10pt}{10pt}38.8 - 38.6 \% \\
Arm & \rule{-12pt}{10pt}51.2 - 52.7 \% & \rule{-10pt}{10pt}54.2 - 54.7 \% \\
Forearm & \rule{-12pt}{10pt}44.4 - 43.6 \% & \rule{-10pt}{10pt}47.2 - 48.6 \% \\
Hip & \rule{-12pt}{10pt}18.7 - 18.7 \% & \rule{-10pt}{10pt}18.1 - 17.9 \% \\
Leg & \rule{-12pt}{10pt}65.8 - 64.9 \% & \rule{-10pt}{10pt}76.8 - 76.9 \% \\
Foreleg & \rule{-12pt}{10pt}62.1 - 62.6 \% & \rule{-10pt}{10pt}72.4 - 72.0 \% \\ \hline
\end{tabular}
\captionof{table}{Body proportions. DWPose vs Perception Neuron sensors. }\label{tab:proporciones}
\end{center}
\end{minipage} 
\hfill 
\begin{minipage}{2.5cm} 
\begin{center}
\includegraphics[width=2.5cm]{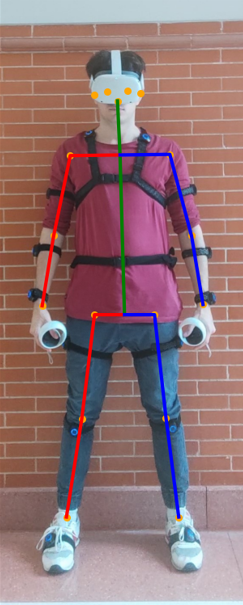}
\captionof{figure}{Keypoints in the image - DWPose}
\label{fig:frame_0}
\end{center}
\end{minipage} 
\end{figure}

Within the scope of this research, we aimed to establish a methodology for acquiring motion metrics that could be standardised between both real-world and virtual environments. Employing a whole-pose estimator \cite{DWPose}, we identify the keypoints of the pedestrian's body in images captured by the front camera of the AV in the real environment (recall Fig. \ref{fig:decision_crossing_event}). Subsequently, the 3D keypoints localised in the virtual environment are projected onto the plane parallel to the crosswalk, aligning with the format of the 2D estimator output. Table \ref{tab:proporciones} outlines the body proportions derived from both procedures for constructing the pedestrian avatar.

Despite the non-correspondence of the keypoints given by Perception Neuron and the 2D estimator, this strategy allows us to conduct an equivalent analysis of the pedestrian's gait from the vehicle's perspective in the two environments. To calculate the pedestrians' pace while crossing, we apply the Fast Fourier Transform (FFT) \cite{FFT} on the lateral position on their ankles, and extract the highest peaks of the frequency spectrum. From the results of the Table \ref{tab:pace_cadence} (Student's t-test cannot be used since it does not involve comparing a series of frequency magnitudes), it can be deduced that the eHMI activation makes pedestrians walk faster ($F_{P}$: t3 $>$ t1 and t4 $>$ t2), while in the VR setup they walk slightly slower in all experimental tasks ($F_{P}$: t$_{i,real}$ $>$ t$_{i,virtual}$). 

The previous inferences are also supported by the count of strides within the 4-second time window defined for the crossing decision (see Table \ref{tab:body_motion_2}). The stabilisation times of both ankles increase when the braking is aggressive or when eHMI is non-activated, indicating that pedestrians halt more their movement in such instances to evaluate the situation, as also shown in Table \ref{tab:body_motion_2}.

\begin{table}[t]
\renewcommand{\arraystretch}{1.2}
\caption{Magnitudes of the three highest peaks, Pace (FFT)}
\begin{center}

\begin{tabular}{ccc|p{1.15cm}p{1.15cm}p{1.15cm}p{1.15cm}}

\multicolumn{3}{c|}{$Crossing$}& \multicolumn{4}{c}{\textbf{Frequencies, Steps per Second}} \\

 \multicolumn{3}{c|}{$pace$ $rate$} & 0.488 Hz & 0.732 Hz & 0.976 Hz & 1.220 Hz \\
\hline

\multirow{9.5}{*}{\rotatebox[origin=c]{90}{\textbf{Task number $i$}}} & \multirow{4.5}{*}{\rotatebox[origin=c]{90}{\textbf{Real setup}} \newline \rotatebox[origin=c]{90}{\textbf{testing}}}   

\rule{-8pt}{12pt} & 1 & 2.304 (f$_3$) & 4.748 (f$_1$) &3.589 (f$_2$) & \\
\rule{-8pt}{10pt} && 2 & 2.570 (f$_3$) & 3.933 (f$_1$) &3.454 (f$_2$) & \\
\rule{-8pt}{10pt} && 3 & & 3.378 (f$_2$) &4.973 (f$_1$) &2.461 (f$_3$) \\
\rule{-8pt}{10pt} && 4 & & 3.188 (f$_2$) &4.372 (f$_1$) &2.817 (f$_3$) \\

\cline{3-7}   
&\multirow{5.5}{*}{\rotatebox[origin=c]{90}{\textbf{Virtual setup}} \newline \rotatebox[origin=c]{90}{\textbf{testing}}}    

\rule{-8pt}{12pt} & 1 & 2.788 (f$_2$) & 3.363 (f$_1$) &2.087 (f$_3$) & \\
\rule{-8pt}{10pt} && 2 & 2.828 (f$_2$) & 3.827 (f$_1$) &2.195 (f$_3$) & \\
\rule{-8pt}{10pt} && 3 & 1.683 (f$_3$) & 4.782 (f$_1$) &2.825 (f$_2$) & \\
\rule{-8pt}{10pt} && 4 & 2.157 (f$_3$) & 4.149 (f$_1$) &2.676 (f$_2$) & \\
                
\end{tabular}
\label{tab:pace_cadence}
\end{center}
\end{table}

\begin{table}[htbp]
\renewcommand{\arraystretch}{1.1}
\caption{Stride count, swing and stabilisation times}
\begin{center}
\begin{tabular}{c|p{0.8cm}p{0.8cm}p{0.8cm}p{0.8cm}} 
\multicolumn{1}{c|}{\emph{4-second window}}& \multicolumn{4}{c}{\textbf{Task number $j$}} \\

 \multicolumn{1}{c|}{\emph{of crossing decision}} & 1 & 2  & 3  & 4 \\

\hline
\rule{0pt}{10pt}Strides (real testing)  & 2.54 & 2.28  & 3.29  & 2.94   \\
SD    & 0.69  & 0.73   & 0.57   & 0.62   \\
Strides (VR testing)      & 2.22  & 2.28   & 2.53   & 2.72   \\
SD      & 0.63 & 0.45  & 0.50  & 0.65 \\
\hline
\rule{0pt}{10pt}Left Swing Phase   & 1.23 s & 1.11 s  & 1.36 s  & 1.26 s  \\
Left Stance Phase    & 2.77 s & 2.89 s  & 2.64 s  & 2.74 s  \\
SD      & 0.36 & 0.30  & 0.25  & 0.31  \\
\rule{0pt}{8pt}Right Swing Phase  & 1.31 s & 1.22 s  & 1.38 s  & 1.24 s  \\
Right Stance Phase   & 2.69 s & 2.78 s  & 2.62 s  & 2.76 s  \\
SD      & 0.33 & 0.30  & 0.22  & 0.28  \\
\end{tabular}
\label{tab:body_motion_2}
\end{center}
\end{table}

\subsection{Subjective Measures}

To make categorical statements regarding the influence of the braking manoeuvre or eHMI on participants' questionnaire responses, we use the Wilcoxon signed-rank test \cite{woolson2007wilcoxon} which is an alternative to Student's t-test when working with ordinal or interval scales. The procedure for this non-parametric statistical test utilised to compare two related samples involves arranging  the values of the absolute differences between the two samples and subsequently calculating a sum of ranks to determine whether the difference between the samples is statistically significant. The null hypothesis of the Wilcoxon test is that there is no difference between the two samples ($H_0:\mu_i\leq\mu_j$), while the alternative hypothesis is that there is a significant difference ($H_1:\mu_i>\mu_j$).

Table \ref{tab:wilcoxon} provides categorical statements, where a 1 in a cell implies rejection of the null hypothesis and acceptance of the alternative hypothesis meaning that the responses to a question in task $i$ (in the row) are significantly greater than those in task $j$ (in the column).

\begin{table}[htbp]
\renewcommand{\arraystretch}{1.2}
\caption{Wilcoxon Signed Rank test, Q1-3, $\alpha$=0.05}
\begin{center}
\begin{tabular}{cccc|p{0.8cm}p{0.8cm}p{0.8cm}p{0.8cm}}
\multicolumn{4}{c|}{\textbf{$H_1:\mu_i>\mu_j$}}& \multicolumn{4}{c}{\textbf{Task number $j$}} \\
 \multicolumn{4}{c|}{} & \colorbox{gray!0}{1} & \colorbox{gray!0}{2} & \colorbox{gray!0}{3} & \colorbox{gray!0}{4}\\
\hline
\multirow{28.5}{*}{\rotatebox[origin=c]{90}{\textbf{Task number $i$}}} 

& \multirow{9}{*}{\rotatebox[origin=c]{90}{\textbf{Q1}}} & \multirow{4.5}{*}{\rotatebox[origin=c]{90}{\textbf{Real setup}} \newline \rotatebox[origin=c]{90}{\textbf{testing}}}                                                    
\rule{0pt}{9pt}& 1   &\colorbox{gray!0}{--}  &\colorbox{gray!20}{1}  &\colorbox{gray!0}{0}  &\colorbox{gray!0}{0} \\
\rule{0pt}{10pt}&&   &2   &\colorbox{gray!0}{0}  &\colorbox{gray!0}{--}  &\colorbox{gray!0}{0}  &\colorbox{gray!0}{0} \\
\rule{0pt}{10pt}&&  &3   &\colorbox{gray!0}{1}    &\colorbox{gray!0}{1}     &\colorbox{gray!0}{--}     & \colorbox{gray!0}{1}\\
\rule{0pt}{10pt}&&  &4   &\colorbox{gray!20}{0}  &\colorbox{gray!0}{1}  &\colorbox{gray!0}{0}  &\colorbox{gray!0}{--} \\

\cline{4-8}
&  &\multirow{4.5}{*}{\rotatebox[origin=c]{90}{\textbf{Virtual setup}} \newline \rotatebox[origin=c]{90}{\textbf{testing}}}                        

\rule{0pt}{9.5pt}&1   &\colorbox{gray!0}{--}   &\colorbox{gray!20}{0}  &\colorbox{gray!0}{0}  &\colorbox{gray!0}{0}  \\
\rule{0pt}{10pt}&& &2   &\colorbox{gray!0}{0}  &\colorbox{gray!0}{--}   &\colorbox{gray!0}{0}  &\colorbox{gray!0}{0}\\
\rule{0pt}{10pt}&&  &3   &\colorbox{gray!0}{1}     &\colorbox{gray!0}{1}     &\colorbox{gray!0}{--}     &\colorbox{gray!0}{1}      \\
\rule{0pt}{10pt}&&  &4   &\colorbox{gray!20}{1}  &\colorbox{gray!0}{1}    &\colorbox{gray!0}{0}     &\colorbox{gray!0}{--}     \\

\cline{3-8}
& \multirow{9}{*}{\rotatebox[origin=c]{90}{\textbf{Q2}}} & \multirow{4.5}{*}{\rotatebox[origin=c]{90}{\textbf{Real setup}} \newline \rotatebox[origin=c]{90}{\textbf{testing}}}                                                    
\rule{0pt}{10pt}& 1   &\colorbox{gray!0}{--}  &\colorbox{gray!0}{0}  &\colorbox{gray!0}{0}  &\colorbox{gray!0}{0} \\
\rule{0pt}{10pt}&&   &2   &\colorbox{gray!0}{1}  &\colorbox{gray!0}{--}  &\colorbox{gray!0}{1}  &\colorbox{gray!20}{0} \\
\rule{0pt}{10pt}&&  &3   &\colorbox{gray!0}{0}    &\colorbox{gray!0}{0}     &\colorbox{gray!0}{--}     & \colorbox{gray!0}{0}\\
\rule{0pt}{10pt}&&  &4   &\colorbox{gray!0}{1}  &\colorbox{gray!0}{0}  &\colorbox{gray!0}{1}  &\colorbox{gray!0}{--} \\

\cline{4-8}
&  &\multirow{4.5}{*}{\rotatebox[origin=c]{90}{\textbf{Virtual setup}} \newline \rotatebox[origin=c]{90}{\textbf{testing}}}                        

\rule{0pt}{10pt}&1   &\colorbox{gray!0}{--}   &\colorbox{gray!0}{0}  &\colorbox{gray!0}{0}  &\colorbox{gray!0}{0}  \\
\rule{0pt}{10pt}&& &2   &\colorbox{gray!0}{1}  &\colorbox{gray!0}{--}   &\colorbox{gray!0}{1}  &\colorbox{gray!20}{1}\\
\rule{0pt}{10pt}&&  &3   &\colorbox{gray!0}{0}     &\colorbox{gray!0}{0}     &\colorbox{gray!0}{--}     &\colorbox{gray!0}{0}      \\
\rule{0pt}{10pt}&&  &4   &\colorbox{gray!0}{1}  &\colorbox{gray!0}{0}    &\colorbox{gray!0}{1}     &\colorbox{gray!0}{--}     \\

\cline{3-8}
& \multirow{9}{*}{\rotatebox[origin=c]{90}{\textbf{Q3}}} & \multirow{4.5}{*}{\rotatebox[origin=c]{90}{\textbf{Real setup}} \newline \rotatebox[origin=c]{90}{\textbf{testing}}}                                                    
\rule{0pt}{10pt}& 1   &\colorbox{gray!0}{--}  &\colorbox{gray!0}{0}  &\colorbox{gray!0}{0}  &\colorbox{gray!0}{0} \\
\rule{0pt}{10pt}&&   &2   &\colorbox{gray!0}{0}  &\colorbox{gray!0}{--}  &\colorbox{gray!0}{0}  &\colorbox{gray!0}{0} \\
\rule{0pt}{10pt}&&  &3   &\colorbox{gray!0}{1}    &\colorbox{gray!0}{1}     &\colorbox{gray!0}{--}     & \colorbox{gray!0}{0}\\
\rule{0pt}{10pt}&&  &4   &\colorbox{gray!0}{1}  &\colorbox{gray!0}{1}  &\colorbox{gray!0}{0}  &\colorbox{gray!0}{--} \\

\cline{4-8}
&  &\multirow{4.5}{*}{\rotatebox[origin=c]{90}{\textbf{Virtual setup}} \newline \rotatebox[origin=c]{90}{\textbf{testing}}}                        

\rule{0pt}{10pt}&1   &\colorbox{gray!0}{--}   &\colorbox{gray!0}{0}  &\colorbox{gray!0}{0}  &\colorbox{gray!0}{0}  \\
\rule{0pt}{10pt}&& &2   &\colorbox{gray!0}{0}  &\colorbox{gray!0}{--}   &\colorbox{gray!0}{0}  &\colorbox{gray!0}{0}\\
\rule{0pt}{10pt}&&  &3   &\colorbox{gray!0}{1}     &\colorbox{gray!0}{1}     &\colorbox{gray!0}{--}     &\colorbox{gray!0}{0}      \\
\rule{0pt}{10pt}&&  &4   &\colorbox{gray!0}{1}  &\colorbox{gray!0}{1}    &\colorbox{gray!0}{0}     &\colorbox{gray!0}{--}     \\
                
\end{tabular}
\label{tab:wilcoxon}
\end{center}
\end{table}

With a confidence level of 95\%, we assert that activating the eHMI enhances the pedestrian's perception of safety (Q1: t3 $>$ t1 and t4 $>$ t2). On the other hand, the smooth braking manoeuvre also increases the feeling of safety, although it is an effect that is not perceived within the virtual setup when the eHMI is disabled (Q1: t3 $>$ t4 and t1 vs t2). Participants appreciate the difference between the smooth and aggressive type of maneuver (Q2: t2 $>$ t1, t3 and t4 $>$ t1, t3). It is worth noting that in the virtual setup the non-activation of the eHMI makes the same manoeuvre feel even more aggressive (Q2: t2 vs t4). Lastly, eHMI is considered to be useful (Q3: t3 $>$ t1, t2 and t4 $>$ t1, t2).

Table \ref{tab:wilcoxon_gap} presents direct comparisons between the questionnaire responses collected from the two setups. Pedestrians feel less safe in the virtual setup when the AV does not communicate its intentions explicitly (Q1: t1$_{real}$ $>$ t1$_{virtual}$ and t2$_{real}$ $>$ t2$_{virtual}$). In addition, they suggest the virtual eHMI has a more positive impact on their decision-making process (Q3: t3$_{virtual}$ $>$ t3$_{real}$).

\begin{table}[htbp]
\renewcommand{\arraystretch}{1.1}
\captionsetup{skip=-20pt}
\caption{Certainty of the discrepancy, Wilcoxon Signed Rank test}
\begin{center}
\begin{tabular}{cc|p{0.8cm}p{0.8cm}p{0.8cm}p{0.8cm}}
 \multicolumn{2}{c|}{Alternative hypothesis:} & \textbf{Task 1} & \textbf{Task 2} & \textbf{Task 3} & \textbf{Task 4}\\
\cline{1-6}
\multirow{2.25}{*}{\rotatebox[origin=c]{90}{\textbf{Q1}} }

    \rule{-10pt}{10pt}&$\mu_{virtual}>\mu_{real}$ & \colorbox{gray!0}{0.6 \%}& \colorbox{gray!0}{7.8 \%}& \colorbox{gray!0}{44.4 \%}& \colorbox{gray!0}{71.9 \%} \\
    \rule{-10pt}{0pt}&$\mu_{real}>\mu_{virtual}$ & \colorbox{gray!20}{99.4 \%}& \colorbox{gray!10}{92.2 \%}& \colorbox{gray!0}{55.6 \%}& \colorbox{gray!0}{28.1 \%} \\ 
    
\cline{2-6}
\multirow{2.25}{*}{\rotatebox[origin=c]{90}{\textbf{Q2}} }

    \rule{-10pt}{10pt}&$\mu_{virtual}>\mu_{real}$ & \colorbox{gray!0}{67.7 \%}& \colorbox{gray!0}{87.9 \%}& \colorbox{gray!0}{58.3 \%}& \colorbox{gray!0}{30.1 \%} \\
    \rule{-10pt}{0pt}&$\mu_{real}>\mu_{virtual}$ & \colorbox{gray!0}{32.3 \%}& \colorbox{gray!0}{12.1 \%}& \colorbox{gray!0}{41.7 \%}& \colorbox{gray!0}{69.8 \%} \\   
    
\cline{2-6}
\multirow{2.25}{*}{\rotatebox[origin=c]{90}{\textbf{Q3}} }

    \rule{-10pt}{10pt}&$\mu_{virtual}>\mu_{real}$ & \colorbox{gray!0}{0.0 \%}& \colorbox{gray!0}{0.0 \%}& \colorbox{gray!20}{95.0 \%}& \colorbox{gray!0}{77.6 \%} \\
    \rule{-10pt}{0pt}&$\mu_{real}>\mu_{virtual}$ & \colorbox{gray!0}{0.0 \%}& \colorbox{gray!0}{0.0 \%}& \colorbox{gray!0}{4.9 \%}& \colorbox{gray!0}{22.4 \%} \\

\end{tabular}
\label{tab:wilcoxon_gap}
\end{center}
\end{table}

Assessing the sense of presence during the VR experiment can help to uncover the reasons of discrepancies in pedestrian crossing behaviour between the real and virtual testing setup. \emph{Self-presence} measures how much users project their identity into a virtual world through an avatar, while \emph{autonomous vehicle} and \emph{environmental presence} examine how users interact with mediated entities and environments as if they were real. Most of the participants perceived the avatar as an extension of their body (M = 4.04, SD = 0.95), including when moving their hands or walking on the road. The vehicle presence was well rated (M= 3.94, SD = 0.97), although not all participants heard the sound of the engine or felt any braking manoeuvre threatening. Environmental-presence (M = 4.34, SD = 0.63) was the most satisfactory, as they claimed to have the feeling of actually being at a crosswalk.

\section{Discussion}

\subsection{Variables Influence in a Real Environment (RQ1)}

Quantitative data shows that participants in the real-world crosswalk experiment notably extended the Space Gap when making their crossing decision if the AV performed a smooth braking manoeuvre. On the contrary, the impact of the "eHMI" factor seemed evident solely when activated alongside gentle braking. Activation of the visual interface did not accelerate pedestrian crossing decisions in instances of aggressive braking manoeuvres. Nevertheless, in the questionnaires, they indicated that both a braking manoeuvre signalling the vehicle's intention to yield and the activation of the eHMI conveyed a greater sense of safety compared to the opposite scenario. The FFT also notes that explicit communication encouraged them to cross the road faster after entering in the lane, while leading to a decrease in eye contact with the AV.

\subsection{Variables Influence in a Virtual Environment (RQ2)}

In the virtual crosswalk experiment, the results reveal that both the smooth braking manoeuvre and the eHMI activation widen the Space Gap when pedestrians decide to cross. In the questionnaires, they report feeling safer when the eHMI is active compared to when it is not, and express a preference for smooth over aggressive braking, but only when the eHMI is operational. Explicit communication results in participants spending less time making eye contact with the AV to assess hazards. Additionally, according to FFT, it prompts them to walk slightly faster once they have entered the lane.

\subsection{Measuring the Behavioural Gap (RQ3)}

A first point to note is that the Student's t-test shows that the space gap L is significantly higher in the real environment than in the virtual environment when the visual interface (i.e., eHMI) is disabled. This finding is supported by the CITs, as pedestrians who spend more time observing the approaching vehicle encounter a smaller space gap L when they eventually decide to cross. Still, we must mention that this discrepancy in the crossing behaviour between the real and virtual testing setup disappears when the eHMI starts working. The CITs and the distance separating the pedestrian from the AV when deciding to cross then are not noticeably different.

The responses to the questionnaire follow the same line of argument. Participants perceive a greater sense of safety in the real environment compared to the virtual environment when the eHMI is deactivated, and feel equally safe when it is activated. This leads us to think the eHMI contributes to increased confidence in the experiment and prompts participants to make the decision to cross earlier. Furthermore, this effect is particularly pronounced in the virtual environment, where the eHMI is most prominently visible, as reported by Q3. Not activating the eHMI heightens the perception of the virtual AV's aggressive braking as even more aggressive (Q2: t2$_{virtual}$ $>$ t4$_{virtual}$). 

\begin{table}[tbp]
\renewcommand{\arraystretch}{1.1}
\captionsetup{skip=-20pt}
\caption{Certainty of the Behavioural Gap, Fisher's method}
\begin{center}
\begin{tabular}{cc|p{0.8cm}p{0.8cm}p{0.8cm}p{0.8cm}}
 \multicolumn{2}{c|}{$H_{1}: L\downarrow, Q1\downarrow, Q2\uparrow, CIT\uparrow$} & \textbf{Task 1} & \textbf{Task 2} & \textbf{Task 3} & \textbf{Task 4}\\
\cline{2-6}
\multirow{6.5}{*}{}

    \rule{-15pt}{10pt}&\emph{More caution in virtual world}& \colorbox{gray!20}{99.9 \%}& \colorbox{gray!20}{99.7 \%}& \colorbox{gray!0}{72.5 \%}& \colorbox{gray!0}{26.4 \%} \\
    \rule{-15pt}{0pt}&\emph{More caution in real world}& \colorbox{gray!0}{0.1 \%}& \colorbox{gray!0}{0.0 \%}& \colorbox{gray!0}{8.5 \%}& \colorbox{gray!0}{52.7 \%} \\

\end{tabular}
\label{tab:fisher_gap}
\end{center}
\end{table}

To gather the evidence on the existence of the \emph{behavioural gap} we employ the Fisher's method \cite{Fisher}, a statistical technique utilised to combine the results of independent significance tests performed on the same data set. The Fisher's method is particularly useful when multiple hypothesis tests are performed and it is desired to combine the evidence from all of these tests to reach an overall conclusion. In Table \ref{tab:fisher_gap} the general alternative hypothesis (H${_1}$) is defined as follows: pedestrians adopt a more cautious crossing behaviour in the virtual world than in the real world (i.e., less Space Gap, less trust Q1, more perceived aggressiveness Q2, more CIT). It is shown that participants demonstrate increased caution in the simulated scenario when the eHMI is inactive, while no conclusive findings can be drawn in the opposite direction.

Comparing the impact of each variable, in the real-world environment an implicit communication is obeyed before an explicit one (Space gap: t1$_{real}$ $>$ t4$_{real}$), while in the virtual environment more trust is placed in explicit communication (Q1: t4$_{virtual}$ $>$ t1$_{virtual}$). FFTs indicate that participants walked more slowly on the road in the virtual environment, probably because they were more curious and entertained by observing the AV, as shown by the eye gazing data.

\subsection{Limitations}

This research was conducted in a simple traffic scenario, featuring only one approaching vehicle, devoid of any social activities in the background. This could have led to collecting information only on individual decision-making and crossing behaviours without the influence of other co-located pedestrians and vehicles. The lighting and weather conditions were also specific (clear sunny day), and results in different contexts may vary.

The immersive VR system for real agents currently employed \cite{CarlaCHIRA2022, CarlaCHIRA23} relies on Unreal Engine 4 and Windows operating system due to the CARLA build and dependencies unique to Meta Quest 2 for Windows. Due to sensors simulation entails a high computational cost, the scene rendering is limited to 15-20 frames per second, which could affect the participants' sense of presence. Moreover, since most of participants had little to no prior VR experience before the experiment, it remains unclear whether the \emph{behavioural gap} results would have differed had the participants been regular users of virtual reality. Increasing the sample size (N = 18) in future studies would allow for a more comprehensive exploration of potential effects, including gender and age disparities in response to the variables investigated. Nonetheless, despite this limitation, we believe the results and conclusions outlined herein provide valuable insights and lay a foundation for further research in the field of real agent simulation.

As demonstrated, the investigation of the \emph{behavioural gap} is intricately tied to specific contextual factors such as the type of scenario, traffic and environmental conditions, etc. Therefore, results cannot be readily generalised across other contexts. However, the methodology presented in terms of combined analysis based on self-reporting and direct measures of behaviour in equivalent real-world and virtual settings, is transferable to other types of scenarios, including different application domains (e.g., robotics). Studying the \emph{behavioural gap} is essential for validating any behavioural data from real subjects interacting with autonomous systems obtained in virtual environments.
\section{Conclusions and Future Work}

This study advances our understanding of the gap between simulation and reality in contexts that incorporate the activity of real agents for autonomous driving research. The digital twin of a crosswalk and an AV was crafted by replicating its driving style and the design of the eHMI it featured, within the CARLA simulator. The participants, who had no previous experience in VR, acted more cautiously in their role as a pedestrian in the simulation by delaying lane entry, slowing their movements and paying more attention to all elements of the environment. This did not prevent us from corroborating the impact of implicit and explicit vehicle communication on the crossing behaviour of pedestrians introduced into the virtual environment. Based on our findings, participants prioritised implicit communication over explicit communication in the real-world scenario, whereas in the VR tests, their decisions were more influenced by explicit communication. 

For future work in this field, we emphasise the importance of familiarising the participants with the VR environment, not only by proposing them to explore the virtual world for a few minutes before starting the tests, but also by involving them in simulated examples with vehicular traffic and street crossings that do not count for the drawing of conclusions. In order to resemble the effects of the braking manoeuvre and eHMI in the simulator to those in the real-world, techniques could be implemented to enhance the \emph{AV presence} rating through more realistic motion dynamics and an engine sound that commensurate with its revolutions. In addition, the brightness of the virtual eHMI could be adjusted to match its showiness in the real environment. If sufficient data were available, a more automatic approach to assessing \emph{behavioural gap} could be achieved, e.g., by learning behavioural differences within a particular scenario and subsequently generating corresponding scores or distances.

\section*{Acknowledgment}
We would like to express our sincere thanks to all participants in the study. 

\section*{Funding details}
This work was funded by Research Grants PID2020-114924RB-I00 and PDC2021-121324-I00 (Spanish Ministry of Science and
Innovation).  S. Martín Serrano acknowledges funding
from the University of Alcalá (FPI-UAH). D. Fernández Llorca acknowledges funding
from the HUMAINT project by the Directorate-General
Joint Research Centre of the European Commission. 

\section*{Disclaimer}
The views expressed in this article are purely those of the authors and may not, under any circumstances, be regarded as an official position of the European Commission.

\section*{Disclosure statement}
The authors report there are no competing interests to declare.

\section*{Safety and Ethical Considerations}
The fundamental pillar guiding the design of the experiments has been the safety and comfort of all participants above any other consideration. On one hand, we chose to implement Level 3 automation in our real testing conditions, despite the fact that Level 4 automation could have been possible. This decision necessitated the presence of a backup driver ready to resume control when needed. In addition, a human supervisor in the rear seats was monitoring the status of all perception and control systems, including access to an emergency stop function. Therefore, human intervention was always possible, both by the backup driver and the supervisor. On the other hand, the braking profile was designed to be extremely conservative, maintaining a substantial margin for reaction, prioritising safety above all else.
Furthermore, we rigorously followed internal and institutional ethical assessment and validation procedures, which included informing the participants and obtaining their written consent, ensuring data privacy, allowing subjects to withdraw from the experiments at any time, and implementing data anonymisation, among other protocols.

\bibliographystyle{IEEEtran}
\bibliography{IVvr23}

\section*{Appendix A}

\subsection*{Self-presence Scale items}

To what extent did you feel that… (1= not at all – 5 very strongly)

\begin{enumerate}
\item You could move the avatar's hands.
\item The avatar's displacement was your own displacement.
\item The avatar's body was your own body.
\item If something happened to the avatar, it was happening to you.
\item The avatar was you.
\end{enumerate}

\subsection*{Autonomous vehicle presence Scale items} 

To what extent did you feel that… (1= not at all – 5 very strongly)

\begin{enumerate}
\item The vehicle was present.
\item The vehicle dynamics and its movement were natural.
\item The sound of the vehicle helped you to locate it.
\item The vehicle was aware of your presence.
\item The vehicle was real.
\end{enumerate}

\subsection*{Environmental presence Scale items} 

To what extent did you feel that… (1= not at all – 5 very strongly)

\begin{enumerate}
\item You were really in front of a pedestrian crossing.
\item The road signs and traffic lights were real.
\item You really crossed the pedestrian crossing.
\item The urban environment seemed like the real world.
\item It could reach out and touch the objects in the urban environment.
\end{enumerate}

\section*{About the authors}

\textbf{Sergio Martín Serrano} is a PhD student at the University of Alcalá. His research interests are focused on the analysis of Vulnerable Road Users (VRUs) behaviours using Virtual Reality (VR) and autonomous driving simulators, covering predictive perception and human-vehicle interaction.  \\

\textbf{Rubén Izquierdo} is Assitant Professor at the University of Alcalá. His research interest focuses on prediction of vehicles' behaviour,  human-vehicle interaction, and control algorithms for highly automated and cooperative vehicles.\\

\textbf{Iván García Daza} is Associate Professor at the University of Alcalá. He specialises in Deep Reinforcement Learning for decision-making in autonomous vehicles and in the application of Explainable AI to decision-making systems. \\

\textbf{Miguel Ángel Sotelo} is Full Professor at the University of Alcalá. His research interests focus on road users' behaviour understanding and prediction, human-vehicle interaction, and Explainable AI for decision-making in autonomous vehicles.  \\

\textbf{David Fernández Llorca} is Scientific Officer at the European Commission - Joint Research Centre, and Full Professor at the University of Alcalá. His research interests include trustworthy AI for transportation, human-centred autonomous vehicles, predictive perception, and human-vehicle interaction.   \\

\end{document}